# Document Embeddings vs. Keyphrases vs. Terms: An Online Evaluation in Digital Library Recommender Systems

Andrew Collins
ADAPT Centre, School of
Computer Science and Statistics,
Trinity College Dublin, Ireland
ancollin@tcd.ie

Joeran Beel
ADAPT Centre, School of
Computer Science and Statistics,
Trinity College Dublin, Ireland
beelj@tcd.ie

## ABSTRACT

Many recommendation algorithms are available to digital library recommender system operators. The effectiveness of algorithms is largely unreported by way of online evaluation. We compare a standard term-based recommendation approach to two promising approaches for related-article recommendation in digital libraries: document embeddings, and keyphrases. We evaluate the consistency of their performance across multiple scenarios. Through our recommender-as-a-service Mr. DLib, we delivered 33.5M recommendations to users of Sowiport and Jabref over the course of 19 months, from March 2017 to October 2018. The effectiveness of the algorithms differs significantly between Sowiport and Jabref (Wilcoxon rank-sum test; p < 0.05). There is a ~400% difference in effectiveness between the best and worst algorithm in both scenarios separately. The best performing algorithm in Sowiport (terms) is the worst performing in Jabref. The best performing algorithm in Jabref (keyphrases) is 70% worse in Sowiport, than Sowiport's best algorithm (click-through rate; 0.1% terms, 0.03% keyphrases).

## 1    Introduction

Many recommendation algorithms are available to operators of recommender systems in digital libraries. The effectiveness of algorithms in real-world systems is largely unreported, and online evaluations of recommender algorithm effectiveness are uncommon [8][11]. We have previously reported the effectiveness of content-based filtering algorithms for recommender systems in digital libraries [5][7][6]. In this paper, we evaluate a standard term-based recommendation algorithm and two promising approaches: keyphrases, and document-embeddings. We compare them through an online evaluation. It is expected that algorithm performance will vary in different domains, for example, movie recommendation and news recommendation. To assess the consistency in performance of these algorithms within similar contexts, we evaluate them in two scenarios within the same domain of scholarly-article recommendation. We consider a "scenario" to be all attributes that describe a recommender system's environment, including users and user interfaces [5].

## 2    Related Work

Keyphrases are typically multi-word phrases that describe a concept or subject within a document [19], for example, "recommender systems research". They may be particularly useful for recommendation when single keywords from a document are not specific enough. De Nart and Tasso compare keyphrase-based recommendations to TF-IDF [18]. They evaluate Lucene's MoreLikeThis TF-IDF implementation and their own keyphrase-based system, using a sample of a MovieLens dataset. Their keyphrase-based approach was ~40% more effective than TF-IDF in their offline evaluation (average nDCG; ~0.27 vs ~0.38).

Doc2Vec [17] is a variation of Word2Vec that learns vector representations of sentences, paragraphs and documents. There exist offline evaluations of Doc2Vec as a recommendation approach. Doc2Vec outperforms TF-IDF in one evaluation (Precision@10: 8.96% vs 62.74%) [12]. The authors conclude that, on their dataset, dense vector representations are better able to capture semantic-similarity between documents for recommendation purposes when compared to simple term-based approaches.

There are few online evaluations that compare multiple algorithms across more than one scenario. Authors have performed "near-to-online" evaluations of the same algorithms across different scenarios [5]. These recommendations are made within the same domain of online news media. It might be expected that a well-performing algorithm on one news website would perform similarly on a different news website. Algorithm effectiveness varied inconsistently, however. For example, the best performing algorithm on one website (cio.de) was the worst performing algorithm on another (ksta.da) ('Most Popular' recommendations; Precision: 0.56 vs 0.01).

We are aware of one comparison of term-based and document-embedding-based recommendation algorithms that uses an online evaluation [12]. The authors also perform a preliminary offline evaluation of both. They use Lucene/ElasticSearch's MoreLikeThis module for TF-IDF, and re-rank its recommendations using Doc2Vec for a test-group. They use click-through rates to compare the algorithms on a large job advertisement website, using 2000 users, over a period of 20 days. The results of their online evaluation are consistent with their offline evaluation: Doc2Vec-re-ranked recommendations achieve an 8% higher click-through rate compared to a standard TF-IDF control.

There are no large-scale online evaluations that compare term, document-embedding, and keyphrase-based recommendation algorithms across multiple real-world scenarios, to the best of our knowledge.

## 3 Methodology

To evaluate the effectiveness of keyphrase, document embedding, and term-based approaches across recommendation scenarios, we examine data from Mr. DLib. Mr. DLib is our recommendation-as-a-service that delivers related-article recommendations to partners [4]. In this work we analyse data from two partners: the digital library *Sowiport*, and the open-source reference manager *Jabref* [16]. Both partners display related-article recommendations from Mr. DLib to their users and represent similar but distinct recommendation scenarios.

Sowiport was the largest digital library for social sciences in Germany [14]. Mr. DLib makes recommendations to Sowiport users from a corpus of 10M articles, and makes recommendations to Jabref users from the Sowiport corpus, and from 14M CORE[1] documents [15]. Related-article recommendations are based on text and metadata from documents in these corpuses, including the title, keywords, and abstract.

A maximum of six recommendations are shown to users at a time in both scenarios. For users of Sowiport, Mr. DLib generates recommendations when a user views an article on the Sowiport website. Related-articles are then shown as a vertical list on the left-hand side of the querying article's detail page (**Fig. 1**). Conversely, JabRef users elect to see recommendations by clicking the "Related Articles" tab within the application (**Fig. 1**). This presents a set of related-articles that Mr. DLib has generated for the currently selected reference in the user's library.

We delivered recommendations over a 19-month period, from March 2017 to October 2018. We primarily delivered term-based recommendations during this time. We did this to provide the best experience to our partner's users, according to our previous evaluations [7]. 95% of recommendations used a term-based algorithm. The remaining 5% of recommendations used our experimental algorithms (3% keyphrases, 2% document-embeddings). All data used is available in the RARD II dataset [10].

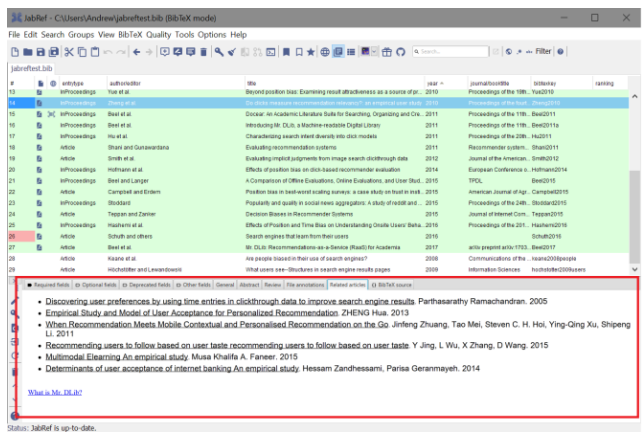

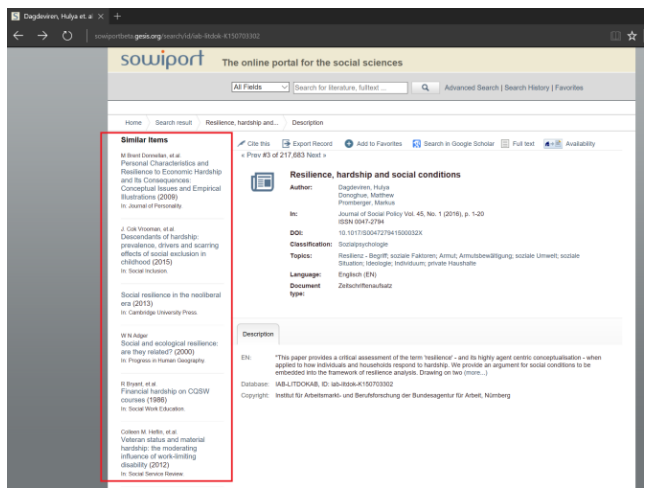

**Fig. 1. Recommendations highlighted in Jabref (top) and for Sowiport (bottom). Users of Jabref choose to see recommendations by clicking the 'Related Articles' tab. Recommendations are always shown to Sowiport users on each item's detail page.**

We measure each algorithm's effectiveness using click-through rate (**CTR**). This is the ratio of clicked recommendations to delivered recommendations. For example, if 1000 recommendations are delivered and 9 are clicked, this gives a CTR of $\frac{9}{1000} = 0.9\%$. CTR correlates with user satisfaction of recommendations [9]. We assume that if an algorithm is effective then, on average, users will interact with recommendations from this algorithm more frequently than recommendations from a less effective one. This will manifest as a higher CTR for the more effective algorithm.

## 3.1 Recommendation algorithms

**Terms**

For term-based recommendations we used Apache Lucene[2]. Mr. DLib first indexes the title and abstract data for all documents in our corpuses. Recommendations are generated for a querying document using Lucene's MoreLikeThis class, which is an implementation of TF-IDF. We used Lucene version 6.3.0 with default MoreLikeThis parameters, and the default set of stop-words[3].

**Document Embeddings**

We use Gensim's[4] implementation of Doc2Vec for our document-embedding-based recommender. We first learn vector-representations of titles and abstracts for all documents in our corpus. Document embeddings are more accurate for titles and abstracts, than titles alone, for finding similar academic texts [2]. To make recommendations, we rank documents according to the cosine similarity between a querying document and candidate documents in our corpus.

**Keyphrases**

Our keyphrase-based recommender is an implementation of work described by Ferrara et al. [13]. We first extract candidate-keyphrases using the Distiller Framework [3]. The title and abstracts of each document are tokenized, part-of-speech tagged, and stemmed. Combinations of keyphrases, that match Distiller's default part-of-speech patterns, are selected. These selected keyphrases are scored according to the statistical properties described by Ferrara et al. [13], for example: the position in the document that the keyphrase was extracted from (depth), or the distance between the first occurrence of the keyphrase and the last occurrence (lifespan). We rank keyphrases according to their overall score, and discard those that are ranked twentieth or worse. We generate unigram, bigram, and trigram keyphrases. We then index each document's keyphrases in Lucene. We also index combinations of n-grams, for example, unigrams and bigrams, bigrams and trigrams, and so on. Our process of keyphrase-selection is illustrated in Fig. 2.

To make recommendations to users based on keyphrases, we randomly choose the type of n-grams to use, for example, unigrams alone, or unigrams and trigrams. We then search the relevant indexed column, using the querying document's keyphrases.

To further evaluate the keyphrase-based recommender, we vary the number of keyphrases used in searches. Approximately 50% of keyphrase-based recommendations were made using a specific, random number of keyphrases (1-19) rather than the maximum available.

**Fig. 2. An illustration of the keyphrase-generation procedure: a) document title and abstracts are tokenized, POS-tagged[5], stopwords are removed, and remaining keywords are stemmed. b) candidate unigram, bigram, and trigram keyphrases are generated. c) candidate keyphrases are weighted and scored [13]**

| a | Title: | | | Research Paper Recommender System - A Quantitative Study of Performance | | | | |
|---|---|---|---|---|---|---|---|---|
| **Tokenisation** | Research | Paper | Recommender | System | A | Quantitative | Study | Of | Performance |
| **POS tagging** | research \ NN | paper \ NN | recommender \ NN | system \ NNS | a \ DT | quantitative \ JJ | study \ NN | of \ IN | performance \ NN |
| **Stopword removal** | research \ NN | paper \ NN | recommender \ NN | system \ NNS | | quantitative \ JJ | study \ NN | | performance \ NN |
| **Stemming** | research \ NN | paper \ NN | recommend \ NN | system \ NNS | | quantiti \ JJ | studi \ NN | | perform \ NN |

| b | Keyphrases | | |
|---|---|---|---|
| | **Unigram** | **Bigram** | **Trigram** |
| | research | research paper | research paper recommend |
| | paper | recommend system | paper recommend system |
| | recommend | paper recommend | system recommend studi |
| | perform | quantit studi | - |

| c | Feature | Weight |
|---|---|---|
| | Depth | 0% |
| | Height | 0% |
| | Lifespan | 0% |
| | Frequency | 10% |
| | Noun Value | 60% |
| | Maximality | 40% |

## 4 Results

Mr. DLib delivered 33.5M keyphrase, document-embedding, and term-based recommendations to Sowiport and Jabref users between March 2017 and October 2018. There were 33,089 clicks on these delivered recommendations, giving an overall click-through rate of 0.1%. The average click-through rate for all algorithms together is 54% lower for Sowiport (0.1%) than Jabref (0.22%). There is a significant difference in the effectiveness of each algorithm between both scenarios (Wilcoxon rank-sum test; $p < 0.05$).

For Sowiport, term-based recommendations outperformed all other algorithms (CTR; 0.1% vs 0.03% for the next best algorithm, keyphrases) (**Table 1**, **Fig. 3**). Document embeddings is the worst performing algorithm for Sowiport (CTR; 0.02%). We find the opposite for Jabref; the term-based approach was less effective than all other algorithms overall (CTR; 0.21%). The most effective algorithm for Jabref is keyphrases (CTR; 0.37%), followed by document embeddings (CTR; 0.25%).

**Table 1. Click-through rates for document embedding, keyphrase, and term-based recommenders, for Sowiport and Jabref. Also shown are the click-through rates for combinations of n-gram types for the keyphrase-based recommender. The best class of algorithm in each scenario is bolded, and the worst algorithm is italicized**

| Algorithm | Jabref | Sowiport |
|---|---|---|
| **Document Embeddings** | 0.25% | *0.02%* |
| **Keyphrases (overall)** | **0.37%** | 0.03% |
|     Keyphrases (unigrams) | 0.33% | 0.04% |
|     Keyphrases (bigrams) | 0.23% | 0.04% |
|     Keyphrases (unigrams and bigrams) | 0.16% | 0.03% |
|     Keyphrases (unigrams and trigrams) | 0.28% | 0.04% |
|     Keyphrases (unigrams, bigrams and trigrams) | 0.26% | 0.03% |
|     Keyphrases (bigrams and trigrams) | 0.62% | 0.03% |
|     Keyphrases (trigrams) | 0.75% | 0.03% |
| **Terms (Lucene)** | *0.21%* | **0.10%** |

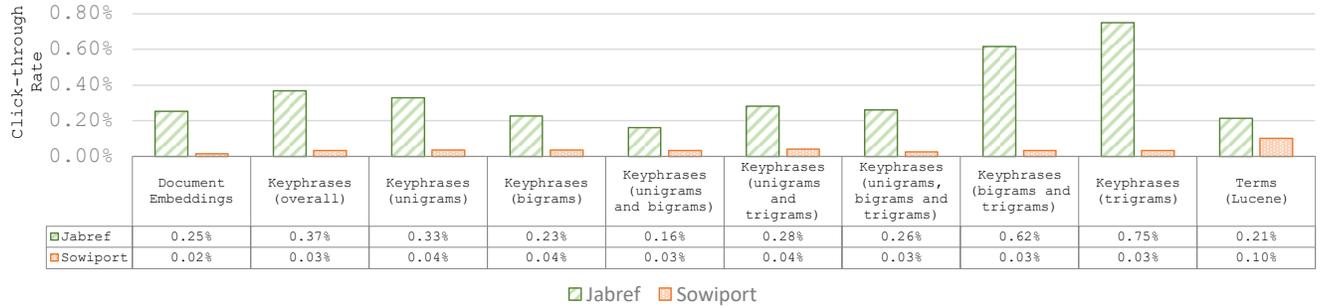

**Fig. 3. Click-through rates for document embedding, keyphrase, and term-based recommenders, for Sowiport and Jabref**

Varying n-gram types for keyphrases is more effective for Jabref than for Sowiport (**Table 1**, **Fig. 3**). Trigram keyphrases are 127% more effective than unigram keyphrases for Jabref, and 368% more effective than unigrams and bigrams together. There is little difference in effectiveness between n-gram types for Sowiport.

Most keyphrase-based recommendations were made using 1-3 keyphrases (**Fig. 4**). Keyphrase count influences effectiveness; for example, recommendations using 8-11 keyphrases had a ~50% lower CTR than recommendations made using 4-7 keyphrases. However, recommendations using more than 12 keyphrases showed an increased CTR compared to 8-11 (CTR; 0.064% vs 0.033%). It is not possible to select an optimal number of keyphrases to use, based on this analysis.

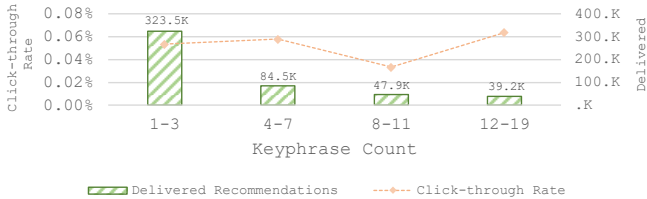

**Fig. 4. The effect of keyphrase count on click-through rates[6]**

## 5 Discussion and Conclusion

Our findings contrast both previous offline and online evaluations of Apache Lucene's TF-IDF. This algorithm has been used as a simple baseline against which all other algorithms tend to show improvements [12, 18]. In our case, TF-IDF is more effective than more recent algorithms for Sowiport. Furthermore, TF-IDF requires less effort to implement, and is theoretically simpler than the other algorithms we have deployed [1].

The range of effectiveness of algorithms is large for both Sowiport and Jabref. We found a ~400% difference in effectiveness between the best and worst algorithm in each scenario. Furthermore, the rank in performance of all algorithms is approximately opposite for Sowiport and Jabref. An algorithm that is useful in one scenario may be almost useless in another, in terms of effectiveness. In this work we analysed just three algorithms across two scenarios. With more algorithms and more scenarios, it is likely that effectiveness will vary even further. Choosing a recommendation algorithm without online evaluation, although the algorithm may seem appropriate, could impact user engagement greatly.

It is not clear why there is such a variation in algorithm performance. Users of Sowiport do not manually choose to see recommendations, whereas Jabref users do. Sowiport users interact with recommendations via a web-browser, whereas Jabref interact via an application for Windows/OSX/Linux. These differences in scenario may influence recommender algorithm effectiveness.

Our evaluation shows that, even within the same domain and across similar recommendation scenarios, effectiveness is inconsistent. These results highlight the need for digital library recommender system operators to evaluate algorithms in an online setting.

## ACKNOWLEDGMENTS

This publication has emanated from research conducted with the financial support of Science Foundation Ireland (SFI) under Grant Number 13/RC/2106. We are grateful to Siddharth Dinesh and Fabian Richter for their work during their internships.

## REFERENCES

[1] Aizawa, A. 2003. An information-theoretic perspective of tf–idf measures. *Information Processing & Management*. 39, 1 (2003), 45–65.

[2] Alvarez, J.E. and Bast, H. 2017. *A review of word embedding and document similarity algorithms applied to academic text*. University OF Freiburg.

[3] Basaldella, M., De Nart, D. and Tasso, C. 2015. Introducing Distiller: a unifying framework for Knowledge Extraction. *Proceedings of 1st AI*IA Workshop on Intelligent Techniques At Libraries and Archives co-located with XIV Conference of the Italian Association for Artificial Intelligence (AI*IA 2015)* (2015).

[4] Beel, J., Aizawa, A., Breitinger, C. and Gipp, B. 2017. Mr. DLib: Recommendations-as-a-service (RaaS) for Academia. *Proceedings of the 17th ACM/IEEE Joint Conference on Digital Libraries* (Toronto, Ontario, Canada, 2017), 313–314.

[5] Beel, J., Breitinger, C., Langer, S., Lommatzsch, A. and Gipp, B. 2016. Towards reproducibility in recommender-systems research. *User modeling and user-adapted interaction*. 26, 1 (2016), 69–101.

[6] Beel, J., Collins, A., Kopp, O., Dietz, L. and Knoth, P. 2018. Mr. DLib's Living Lab for Scholarly Recommendations. *arXiv preprint arXiv:1807.07298*. (2018).

---

[6] Not all documents have as many as 19 keyphrases available. Fewer recommendations can be made using 19 keyphrases than with 1 keyphrase, for example.

[7] Beel, J., Dinesh, S., Mayr, P., Carevic, Z. and Raghvendra, J. 2017. Stereotype and Most-Popular Recommendationsin the Digital Library Sowiport. (2017).

[8] Beel, J., Gipp, B., Langer, S. and Breitinger, C. 2016. Research-paper recommender systems: a literature survey. *International Journal on Digital Libraries*. 17, 4 (Nov. 2016), 305–338.

[9] Beel, J. and Langer, S. 2015. A Comparison of Offline Evaluations, Online Evaluations, and User Studies in the Context of Research-Paper Recommender Systems. *Proceedings of the 19th International Conference on Theory and Practice of Digital Libraries (TPDL)* (2015), 153–168.

[10] Beel, J., Smyth, B. and Collins, A. 2019. RARD II: The 94 Million Related-Article Recommendation Dataset. *Proceedings of the 1st Interdisciplinary Workshop on Algorithm Selection and Meta-Learning in Information Retrieval (AMIR)* (2019).

[11] Champiri, Z.D., Asemi, A. and Binti, S.S.S. 2019. Meta-analysis of evaluation methods and metrics used in context-aware scholarly recommender systems. *Knowledge and Information Systems*. (2019), 1–32.

[12] Elsafty, A., Riedl, M. and Biemann, C. 2018. Document-based Recommender System for Job Postings using Dense Representations. *Proceedings of the 2018 Conference of the North American Chapter of the Association for Computational Linguistics: Human Language Technologies, Volume 3 (Industry Papers)* (2018), 216–224.

[13] Ferrara, F., Pudota, N. and Tasso, C. 2011. A Keyphrase-Based Paper Recommender System. *Digital Libraries and Archives: 7th Italian Research Conference, IRCDL 2011, Pisa, Italy, January 20-21, 2011. Revised Papers*. M. Agosti, F. Esposito, C. Meghini, and N. Orio, eds. Springer Berlin Heidelberg. 14–25.

[14] Hienert, D., Sawitzki, F. and Mayr, P. 2015. Digital library research in action–supporting information retrieval in sowiport. *D-Lib Magazine*. 21, 3/4 (2015).

[15] Knoth, P. and Zdrahal, Z. 2013. CORE: aggregation use cases for open access. *Proceedings of the 13th ACM/IEEE-CS joint conference on Digital libraries* (2013), 441–442.

[16] Kopp, O., Breitenbücher, U. and Müller, T. 2018. CloudRef-Towards Collaborative Reference Management in the Cloud. *ZEUS* (2018), 63–68.

[17] Le, Q. and Mikolov, T. 2014. Distributed representations of sentences and documents. *International Conference on Machine Learning* (2014), 1188–1196.

[18] Nart, D.D. and Tasso, C. 2014. A Personalized Concept-driven Recommender System for Scientific Libraries. *IRCDL* (2014).

[19] Siddiqi, S. and Sharan, A. 2015. Keyword and keyphrase extraction techniques: a literature review. *International Journal of Computer Applications*. 109, 2 (2015).